# Nuclear fusion for AI: A pathway to power data centers sustainably

Layla Araiinejad[1*] and Vineet Jagadeesan Nair[1*]
[1] Massachusetts Institute of Technology, Cambridge, MA 02139, USA
*Corresponding authors (equal contribution): layla00@mit.edu, jvineet9@mit.edu

## Abstract
This perspective examines whether nuclear fusion can provide a scalable, low-carbon power source for rapidly growing AI-driven data center demand. As large language models, cloud computing, and cryptocurrency mining accelerate electricity consumption growth, data centers are projected to account for a substantially larger share of U.S. and global electricity use in the coming decades, creating significant pressure on grid reliability and decarbonization goals. We evaluate the technical and economic alignment between data center load profiles and nuclear power, particularly fusion, through a comparative analysis of capacity factors, levelized cost of electricity, grid interconnection constraints, and deployment pathways. Unlike intermittent renewables, nuclear fission and fusion offer high-capacity-factor, firm baseload generation suited to AI training and inference workloads that require continuous, reliable power. Preliminary techno-economic analysis suggests that several Nth-of-a-kind fusion concepts, particularly magnetic confinement systems, may become cost-competitive with firmed renewable systems and advanced fission for hyperscale data center applications. Co-location of fusion plants with data centers further reduces transmission bottlenecks, improves resilience, and aligns with emerging hyperscaler procurement strategies. We also assess recent regulatory developments and argue that fusion's favorable safety profile and reduced waste burden improve its long-term social and political viability relative to fission. We conclude that fusion represents a strategically important pathway for sustainably powering next-generation computing infrastructure and should be prioritized in both policy and industrial deployment planning.

## Motivation and background

The explosion of cryptocurrency mining, blockchain, cloud computing, artificial intelligence (AI), and machine learning (ML) has led to massive load growth in recent years. Annual US data center electricity demand is estimated to double from around 4% today to 9.1% in 2030[1] (of total US electricity generation), and up to 6% by 2026 [2]. Yet, there is some uncertainty associated with these predictions, with some estimates forecasting data centers to make up as much as 12% of total US electricity consumption by 2028 [3]. Currently, cryptocurrency mining and conventional data centers (along with associated transmission networks) account for the majority of this demand, making up about 2.4-3.3% of global electricity consumption [4] in 2022 and around 1% of global energy-related emissions in 2020 [5]. However, the demand for AI model training and inference is growing rapidly, with 13-27% compounded annual growth rates expected for the next decade. This demand for data center loads will be further magnified due to the growth of cryptocurrencies and increased computing workloads arising from the digitalization of other sectors like transportation, the power grid (with devices like smart thermostats and smart inverters), and manufacturing. Thus, the near- and medium-term growth in data center demand should also be considered alongside the larger-scale, longer-term growth expected due to factors like electric vehicle adoption, onshoring of manufacturing, and increased electrification of buildings and heavy industry.

Large language models (LLMs) have been the dominant driver behind AI electricity demand growth. In addition to several novel theoretical breakthrough in ML techniques [6], another major driving factor behind the impressive capabilities of large language model has been model overparameterization (with billions to trillions of parameters) [7]. Training massive models have also required large datasets and extremely large compute budget, using advanced graphical processing units (GPUs), tensor processing units (TPUs), and application-specific integrated circuits (ASICs). It is estimated that the amount of computation used to train state-of-the-art AI models has increased by roughly 350,000 times since 2014, with the increasing proliferation of generative vision, language, and video models [8]. The main source of electricity consumption (as high as 40%) at data centers is the high cooling needs during operation [1], to maintain servers and chips within safe temperature ranges. While air cooling was largely sufficient for the previous generation of computing chips, the latest advanced chips such as those from NVIDIA require liquid cooling due to the higher heat output and increased power density of such data center racks [9].

The load growth from data centers comes alongside demand increases from the electrification of transportation, heating, and heavy industry. These compounded events are increasingly placing severe stress on the grid. This sudden demand growth bucks the historical trend where US electricity demand had remained relatively flat over the past few decades, owing to energy efficiency gains. As a result, there is an urgent need for rapidly deploying clean energy generation at similar rates to curb greenhouse gas emissions and meet climate change mitigation goals [9]. Sustainably powering data centers is also crucial for tech companies and other hyperscalers to meet their own ambitious internal goals, such as commitments from Google and Microsoft to achieve 100% carbon free electricity for their operations by 2030 and Apple's aim to be carbon-neutral (net-zero emissions) by 2030 [10]. Amazon has already matched 100% of their global energy consumption with electricity from renewable sources [11]. However, the AI boom has caused increases in annual emissions of some of these companies in the past 5 years. For instance, $CO_2$ emissions had risen by nearly 30% since 2020 for Microsoft, and by almost 50% since 2019 for Google, largely due to their data center expansion [12]. Moreover, these climate goals can be even more challenging depending on how they're defined. So far, tech companies have largely relied on a combination of clean power, carbon credits and renewable energy certificates to achieve net-zero emissions overall on a year-end basis. However, there is now greater emphasis on being truly net-zero during operation. As mentioned, one example of this is Google's aim to achieve 24x7 clean energy for their operations for 2030 [13]. This requires matching all electricity consumption in real-time with zero-emissions power, which is much more challenging due to the intermittency, uncertainty and variability associated with renewables like wind and solar [14]. Nuclear power (either fission or fusion) is an attractive alternative to meet such carbon-free matching requirements and ensure data center workloads are always met sustainably, without having to rely on carbon credits or RECs.

## Nuclear energy as a solution

Solar and wind power have relatively low capacity factors (CFs), globally averaging around 21% and 32%, respectively. This is significantly lower than fossil fuel-based generation like coal and gas which offer higher CFs in the range of 40-50% and 40-60%, respectively (depending on the type of plant) [15]. On the other hand, nuclear fission power plants have much higher capacity factors approaching 93%, allowing them to effectively provide firm, baseload zero-carbon power [16]. Although first-of-a-kind (FOAK) fusion plants are expected to have comparatively lower availability and capacity factors (around 75%), Nth-of-a-kind (NOAK) plants will approach fission levels (around 85-90%) in the coming decades [17], allowing it to meet the high uptime requirements of premium tier III and tier IV data centers used for AI training and inference. While renewable sources like geothermal and hydroelectric power do offer high factors closer to 70% and 40%, respectively, these are constrained in terms of where they can be located. Thus, nuclear is potentially a better fit solution since it offers more flexibility in location and be sited closer to the desired data center site. Their baseload characteristic also makes them a better fit for data center owners who prioritize reliability. Achieving a similar level of reliability with renewables requires significant amounts of on-site battery storage in addition to grid connection, which in turn increases total project costs in the form of 'firming costs'. However, one issue with nuclear power is that it generally has high minimum technical output and limited ramping capabilities (lower ramp rates) than fossil fuel sources, due to technical constraints and safety considerations [18]. This makes it less suitable for load following generation, especially compared to gas turbines and hydropower (or pumped storage). However, data center owners and operators do usually have some degree of controllability over their electricity consumption, by modifying their operations in real-time. Thus, as a workaround to deal with slower ramping capabilities of nuclear power plants, data center owners or operators can potentially shape their load profiles to be closer to a baseload. There has also been significant research in the area of data center load flexibility and demand response (DR) [19][20]. By modifying workloads in real-time, data center owners could either ramp up or ramp down their electricity use.

## Technical alignment between data center needs, nuclear fission, and fusion

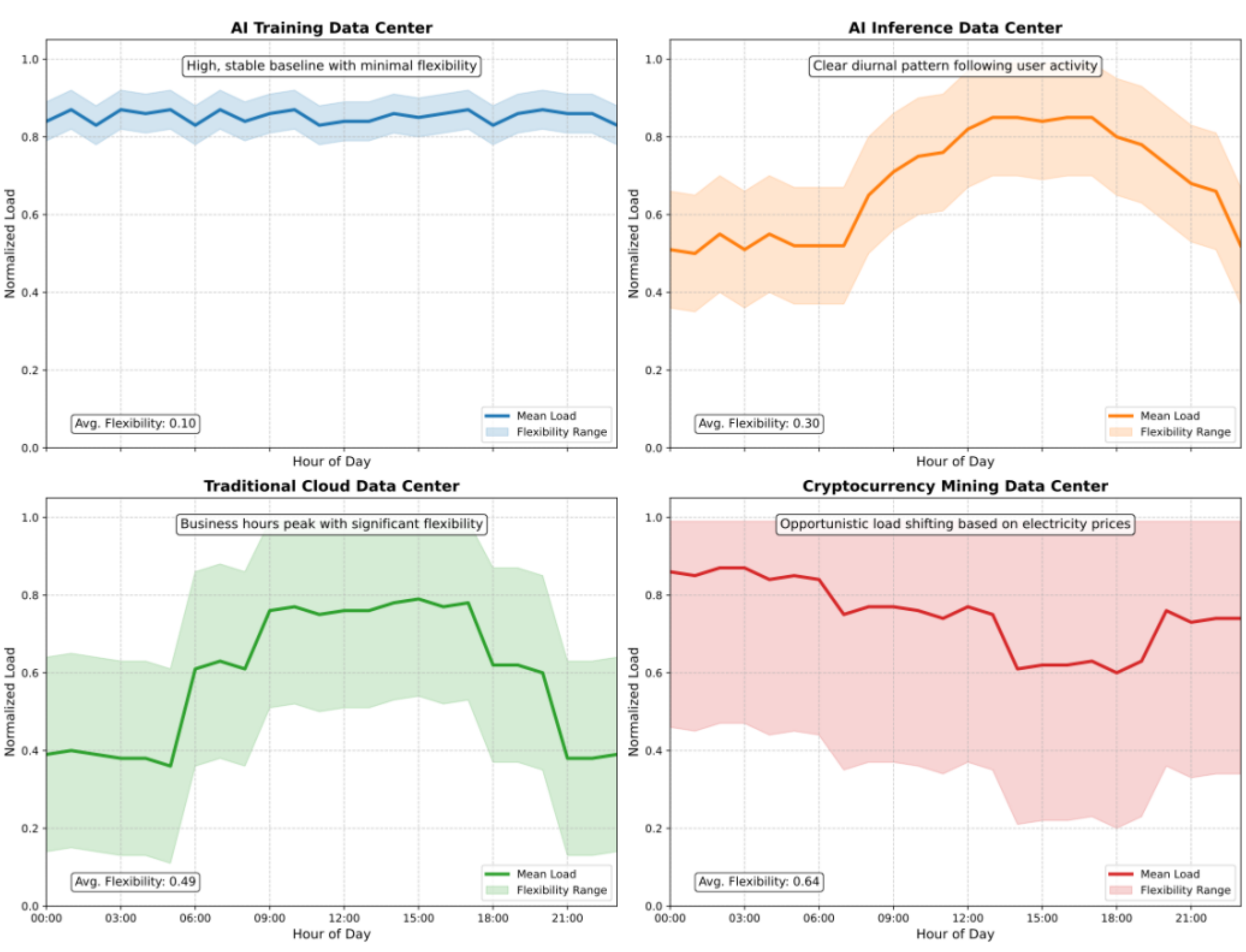

Figure 1: Synthetic examples of representative data center load and flexibility profiles.

However, different types of data centers have different constraints that affect their flexibility characteristics and availability, as seen in Figure 1. 'Conventional' data centers used for traditional cloud computing tasks have moderate flexibility since their non-critical workloads can be shifted in both time and space (across multiple data center hubs) depending on task priorities. These usually display high, steady 24/7 load with some daily variations but the global user base smooths peaks, with recent pilots at Google showing up to 10-20% load curtailment is feasible without adverse service impacts [21]. On the other hand, data centers used for AI and ML model training function like large supercomputers with shared memory across different processors. The model training process, which essentially involves dense linear algebra operations on large matrices, is efficiently parallelized across large numbers of GPUs and TPUs all located at the same data center. This limits the capability to shift training workloads spatially. Temporal flexibility is also limited since this would imply interrupting training. However, there is still some potential to provide temporal shifting by leveraging characteristics of the training process. For instance, pausing and restarting training using model checkpoints is possible, although this is often undesirable and costly for AI companies. There are also sometimes redundant computations involved in massively parallelized training distributed across multiple nodes. This means that some nodes (with duplicate operations on the same data) could potentially drop out to scale down compute without significantly affecting the overall job. Recent industry research projects have studied the potential of carbon-aware load shifting for data centers, such as this work from Google [22]. However, overall, AI training data centers have significantly less flexibility, which makes their load growth even more challenging for the grid [23]. These peak at full capacity for long periods during training rains and require extremely reliable and non-interruptible power supply and thus need to be sited at areas that support continuous heavy load. Scheduling training runs and cooling support systems can provide limited demand response. AI inference training centers are similar to traditional cloud computing, with hourly and daily swings in demand varying with user demand and search queries, but distribution across several sites reduces peaks. They offer limited deferability since requests need to be met in real-time but can be shifted to other data centers at nearby hubs if needed. Finally, cryptocurrency mining sites offer the most flexibility due to the nature of their operation and very simple computational tasks involved. They typically run at peak output 24/7 but can be shut off nearly completely based on market conditions or grid conditions or can also increase load to absorb excess renewables generation. From the grid's perspective, data center load growth can largely be characterized as an increase in steady baseload demand, with relatively limited peak variability and minimal demand-side flexibility. This load profile makes nuclear fission and fusion particularly attractive generation options, as they are well suited to supplying continuous, reliable power at scale. Their high capacity factors and firm generation characteristics are especially advantageous for AI training and inference workloads, which impose stringent reliability and uptime requirements.

## Feasibility and Preliminary Techno-Economic Analysis

Renewable energy sources such as wind and solar are imperative for the clean energy transition but suffer from low power density, inefficient baseload performance, and a lack of cost-effective storage solutions. Nuclear fission, while supplying a substantial share of the world's non-carbon-emitting electricity, faces high capital costs, persistent cost overruns, and regulatory burdens stemming from lengthy permitting processes and stringent safety requirements [24]. Together, these constraints have prevented either technology from fully displacing fossil fuels. Fusion energy represents a potential solution to both. This paper argues for co-locating new generation directly with data center loads—a configuration that allows a reactor to serve dedicated industrial demand without affecting retail electricity prices for consumers. To understand why this matters, it helps to situate fusion within the broader nuclear landscape. The European Commission defines SMRs as having a maximum output of 300 MWe, capable of producing approximately 7.2 million kWh per day, compared to conventional large-scale plants exceeding 1,000 MWe [25]. Large-scale fission in the United States has exhibited a negative learning rate per unit deployed — costs have gone up, not down, with deployment, largely because of regulatory burden. Developers have responded by pivoting to SMRs, though the economics remain unfavorable: according to the 2024 MIT CANES report, the levelized cost of electricity (LCOE) of a 300 MWe water-cooled SMR is approximately 45% higher than the NOAK LCOE for the AP1000. Only two SMRs have been deployed worldwide to date, neither in the United States [26]. The most concrete near-term fission deployment in the U.S. is TerraPower's Natrium reactor. On March 4, 2026, the NRC voted to award Kemmerer Unit 1 in Wyoming a construction permit—the first ever issued to a commercial-scale advanced nuclear plant [27]. Construction begins in 2026, with an earliest operating date of 2030. Whether that schedule holds is another question; U.S. nuclear construction has a well-documented history of delays and cost overruns, and there is little structural reason to expect Natrium will be exempt from those pressures.

The private fusion industry, meanwhile, has attracted serious capital from technology companies facing their own power constraints. In 2023, Microsoft signed a 50 MW PPA with Helion Energy for delivery by 2028—an aggressive target by any measure[28] . Sam Altman had previously invested $375 million in Helion, reportedly with the intention of using fusion to power a data center[29]. The underlying demand is real: Microsoft and OpenAI's Stargate project is reported to cost $100 billion and will require approximately 5 GW of electrical power, with a construction timeline of five to six years [30]. Whether 2028 fusion delivery is achievable to meet that demand is a separate question from whether the demand itself exists—it clearly does. Other developers are targeting the early 2030s with somewhat more conservative timelines. Type One Energy has begun construction of a stellarator fusion plant within a retired coal facility, with grid delivery expected in the early 2030s [31]. Commonwealth Fusion Systems plans to bring its first plant online in Chesterfield County, Virginia on a similar timeline, and has already signed a PPA with Google to supply 200 MW to its data centers [32]. The clustering of these announcements around the early 2030s reflects both where fusion physics currently stands and the commercial pressure to deliver.

Techno-economic analysis provides a framework for situating these developments within a broader cost picture. TEA has informed U.S. energy policy since the 1970s, when the National Energy Modeling System (NEMS) was introduced to project energy production, imports, conversion, consumption, and prices under varying macroeconomic and technology assumptions [33] Figure 2[35]. Similarly, the rationale for public investment in fusion research has long depended on the prospect that fusion could ultimately become an economically competitive source of electricity. As a result, many reactor studies have been conducted not only to demonstrate technical feasibility but also to explore pathways toward economically viable fusion power plants.

A wide range of fusion concepts have been proposed over the years, and these designs continue to evolve as the field advances and private-sector investment accelerates. The most widely pursued approach is magnetic confinement (MC) fusion using a deuterium-tritium (D-T) fuel cycle, although other concepts such as inertial confinement fusion (ICF) and magneto-inertial fusion (MIF) have also received significant attention. Given the diversity of reactor concepts, no single design can be viewed as representative of the eventual commercial fusion industry. Nevertheless, conceptual reactor studies provide a useful basis for economic comparison and techno-economic assessment. Because fusion has not yet reached commercial deployment, there is little empirical information available on the costs of constructing and operating fusion power plants. Consequently, conceptual reactor studies remain the primary source of information for evaluating fusion economics. Some studies directly estimate capital and operating costs, while more recent analyses build upon existing reactor concepts and incorporate advances in materials, engineering design, and regulatory assumptions. For example, Araiinejad and Shirvan developed a techno-economic assessment based on a modified ARC tokamak design, incorporating updated assumptions regarding candidate materials and design parameters drawn from the fusion literature and industry sources [36] . The study provides lower- and upper-bound estimates for the levelized cost of electricity (LCOE) from a D-T magnetic confinement fusion power plant. For the purposes of this comparison, the lower-bound estimate is used, as it reflects a regulatory framework in which fusion maintains a comparatively low radioactive inventory and therefore faces a less burdensome licensing environment than conventional fission technologies.

Cost estimates for magneto-inertial fusion are drawn from a 2017 ARPA-E-sponsored study of four compact, modular fusion concepts [37]. The study estimated average capital costs of $2.4/W and $1.2 billion for plants averaging roughly 500 MWe under NOAK assumptions and also provided cost of electricity (COE) estimates for the four concepts. The reported COE values were adjusted using the Consumer Price Index to express them in current dollars and provide a consistent basis for comparison with the other generation technologies. While future commercial fusion plants may differ substantially from these conceptual designs, such studies provide a valuable basis for understanding the potential economics of fusion energy and comparing alternative technological approaches.
Comparing the cost of electricity estimates for Magnetic Confinement and Magneto-Inertial Fusion with unsubsidized costs from the Lazard 2024 LCOE report suggests that both could fall within the cost range of existing generation technologies under mature deployment assumptions [38]. The wind and solar values in this comparison include firming and storage costs; the fission bounds are derived from Vogtle Units 3 and 4. Two limitations are worth noting: Lazard's firming methodology uses natural gas peakers, meaning the renewable cost estimates are not truly zero-emission; and fusion concepts with higher physics risk but greater potential for capital cost reduction are excluded due to insufficient costing data.

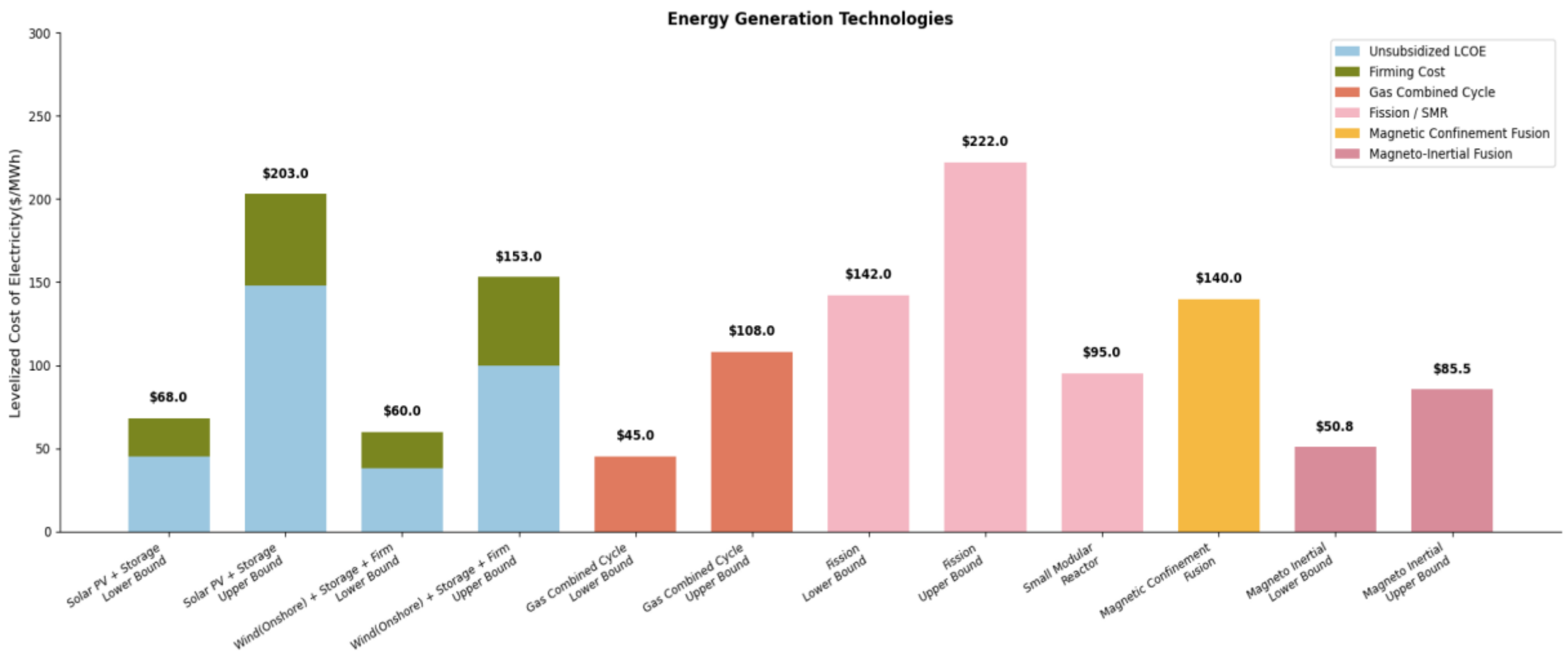

**Figure 2: Comparison of LCOE among top candidate power sources for data centers**

Figure 2 above presents NOAK cost estimates, reflecting mature deployment conditions rather than first-of-a-kind capital costs. For data center applications, the relevant comparison is not the lower bound of renewables but the upper bound — firm, dispatchable power is the requirement, not cheap average electrons. Solar PV with storage reaches $203/MWh at the upper bound; wind with firming reaches $153/MWh. Against those figures, Magnetic Confinement fusion at $140/MWh is already cost-competitive, and even the SMR estimate of $95/MWh sits within that range. Magneto-Inertial Fusion, at $50.8/MWh to $85.5/MWh, falls below the upper-bound costs of firmed solar and wind and overlaps the $45/MWh to $108/MWh range for natural gas combined-cycle generation. Under NOAK assumptions, these estimates suggest that some fusion technologies could supply data center loads at costs comparable to natural gas while avoiding the emissions associated with gas-fired generation. . It bears noting that the fusion figure represents NOAK projections under assumed learning curve and commercial-scale conditions; no fusion power plant has yet been built, so FOAK costs remain speculative. The comparison is nonetheless meaningful for evaluating long-run economic viability — and for the load profile that data centers actually present, several fusion concepts are plausibly within reach at NOAK. A recent Idaho National Laboratory study on powering data centers with clean energy models this requirement at the system level, finding that substantial capacity overbuild is often necessary. In some cases, installed capacity reaches multi-gigawatt levels to maintain reliability during periods of low wind and solar generation [39]

Renewable energy and storage technologies have generally experienced strong positive learning rates (LRs) in terms of lowering capital investment costs and LCOE and increasing capacity factor over time. Utility-scale solar PV has shown the most dramatic and rapid cost declines, while onshore and offshore wind costs have also declined but at slightly slower rates [40]. On the other hand, costs for nuclear fission have either held steady or have even experienced slightly negative learning rates in recent decades, mainly due to regulatory and financing issues that have raised costs and made it very sensitive to the cost of capital [41].

A recent 2026 study found that fusion is expected to have experience average LRs close to 5%, which is higher than the average rate of 2% for fission, but much lower than the LRs of 12%, 20%, and 23% for onshore wind, Lithium-ion batteries, and solar PV, respectively. However, there is significant uncertainty in the cost evolution of fusion power plants – with projected LRs ranging from 2% to 20% for both magnetic and laser-based inertial fusion energy [42]. In terms of availability, while capacity factors for both solar and wind have improved gradually over time, they are still far below fission as well as projected values for fusion [43]. Fission plants have improved their availability from around 50% in 1970 to 80-90% today [17]. Uncertainty around US climate policy, especially at the federal level, also affects the economics of future nuclear projects as well as renewables and storage. Historical investment- and/or production-tax credits have played a significant role in reducing total costs of zero-carbon projects, while carbon credits affect the viability of carbon capture and sequestration (CCS) technologies – which can be used to reduce the emissions associated with data centers power by natural gas. Purely on the basis of LCOE, the US Energy Information Administration has estimated that advanced nuclear (primarily fission) will be cost competitive with offshore wind by 2030 but will be a premium compared to natural gas (with or without CCS) and solar-battery hybrids, while solar PV and onshore wind are projected to be the cheapest options even without tax credits. However, dispatchable sources like nuclear, gas, and batteries more offer value to the grid than non-dispatchable renewables, in terms of their energy, capacity, and spinning reserve contributions. This is captured by the levelized avoided cost of electricity (LACE), as opposed to LCOE which only reflects the cost to build and operate plants [44].

Nuclear fission has often shown cost increases over time rather than declines in the United States. In the United States, reactor costs fell by about 14 percent per year from 1954 to 1968 [45]. That pattern reversed by the late 1960s, with costs rising at roughly 23 percent per year, then continuing to increase at about 5 to 10 percent per year after the Three Mile Island accident [45]. The shift appears structural rather than temporary. Safety requirements tightened. Licensing durations extended. Construction timelines lengthened. These factors compound through higher financing costs and greater exposure to design changes during construction. Standardization did not take hold in the United States. More than 50 reactor designs were deployed, limiting repetition and constraining supply chain learning. Cost reductions are more consistent in factory-based manufacturing with fixed designs and repeated production. They are weaker in large, site-built projects with high variability. Historical learning curves can mislead when used to project future costs under changing conditions. Cost reductions depend more directly on standardization, shorter construction timelines, regulatory alignment with repeat builds, and a shift toward manufacturable components. Thus, the fusion industry can leverage this experience from fission and prioritize standardized designs, modularization, and manufacturing-led deployment pathways early in commercialization.

## Policy and Regulatory Pathways for Fusion

U.S. energy policy is moving in a direction that favors nuclear deployment. U.S. Secretary of Energy Chris Wright has made clear that nuclear power will be actively incentivized domestically, which raises an immediate question about whether existing regulatory and financial instruments are actually adequate, or whether the framework needs reform before fusion can meaningfully participate. The fission buildout, if it proceeds, also carries indirect benefits for fusion that are easy to underestimate. Public familiarity with nuclear energy tends to reduce political resistance to technologies in the same family, and fusion's environmental profile gives it a stronger starting position than fission ever had in that regard.

The most significant regulatory development to date is the Fusion Energy Act of 2024, passed as part of the bipartisan FIRE Grants and Safety Act. The Act amended the Atomic Energy Act of 1954 to formally separate fusion machines from fission reactors and directed the NRC to develop a dedicated, technology-inclusive licensing pathway by 2027 [46]. What makes this notable is the vehicle: this is federal statute, not agency guidance or a policy initiative subject to administrative reversal. Congress made fusion regulation a legal mandate, which matters considerably for investment certainty and for how other countries read the direction of U.S. nuclear policy. The underlying technical rationale is straightforward. Fusion does not sustain a chain reaction, produces no high-level radioactive waste, and presents a risk profile that fission's regulatory architecture was never designed to accommodate. The Act begins to correct a mismatch that should have been addressed earlier.

Most countries either adopt IAEA safety standards, themselves reflecting decades of U.S. and NRC input, or mirror the regulatory approach of whatever country is supplying their reactors. Neither path produces truly independent frameworks, which means U.S. regulatory choices carry significant weight abroad. The United Kingdom moved first on fusion-specific regulation, opting to place fusion facilities under the Health and Safety Executive rather than its nuclear fission regulator, an explicit acknowledgment that the two technologies warrant different treatment[47]. The U.S. is now following with a statutory framework of its own, and given the concentration of private fusion investment in the United States and the NRC's long track record of influencing international practice, the American approach is likely to become the default reference point for countries that lack the capacity to build their own frameworks from scratch. How fusion machines are defined, how jurisdiction is assigned, and how the licensing process is structured in Washington will reverberate well beyond U.S. borders.

There is also a separate but related dynamic in countries where fission has been politically foreclosed. Germany shut down its last reactors in April 2023. [48] Austria has a constitutional ban on fission and never brought its completed reactor online after a 1978 referendum went against it [49]. Italy walked away from fission in 1990 and confirmed that decision after Fukushima. Belgium has extended two reactors through 2035 but remains politically ambivalent about any further nuclear expansion. New Zealand, Denmark, Portugal, Ireland, Norway, Greece, and Australia all maintain legal or political resistance to fission with no operating domestic capacity. In several of these cases the opposition is specifically targeted at fission's risk characteristics: waste, accident potential, the chain reaction itself. Fusion sidesteps each of those objections on technical grounds. Whether that translates into political viability is a different question, and it would be naive to assume anti-nuclear constituencies will automatically distinguish between the two technologies. Public acceptance would need to be built on fusion's own terms. But the opening exists in a way it does not for fission.

The financial picture is where the near-term action is. The data center buildout has created a demand signal that did not exist five years ago. Hyperscale operators are signing PPAs years ahead of delivery at prices that reflect the premium they are willing to pay for firm, carbon-free power. Microsoft and Google have net-zero commitments and cannot wait for the grid to decarbonize on its own timeline [50]. Fission can serve some of that demand, but high-level waste and long-term storage remain genuine liabilities for companies operating under sustained public and investor scrutiny on environmental performance. Fusion's combination of high power density, no long-lived waste, and no chain reaction risk address those liabilities in a way that fission structurally cannot. The policy instruments that would accelerate FOAK deployment are not novel mechanisms. Federal loan guarantees, production and investment tax credits, long-term government anchor PPAs, and co-location incentives are the same tools that moved renewables from demonstration to commercial scale over the past two decades. The commercial pressure behind fusion is real and growing. The question is whether the policy framework moves fast enough to meet it before that demand is absorbed by other technologies.

## Grid impacts and scenarios for deployment

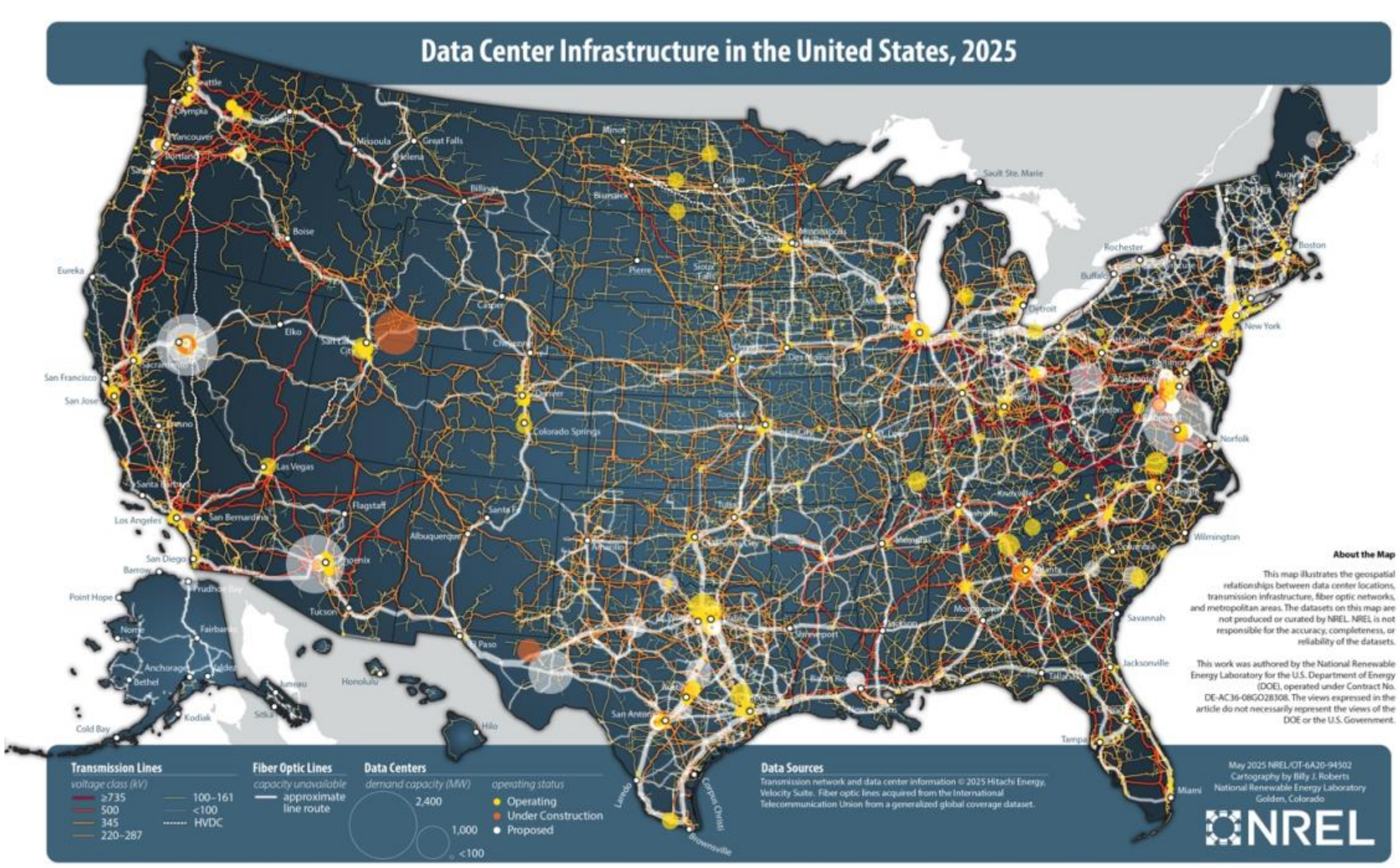

**Figure 3: Data center infrastructure map of the US. Source: National Laboratory of the Rockies [51].**

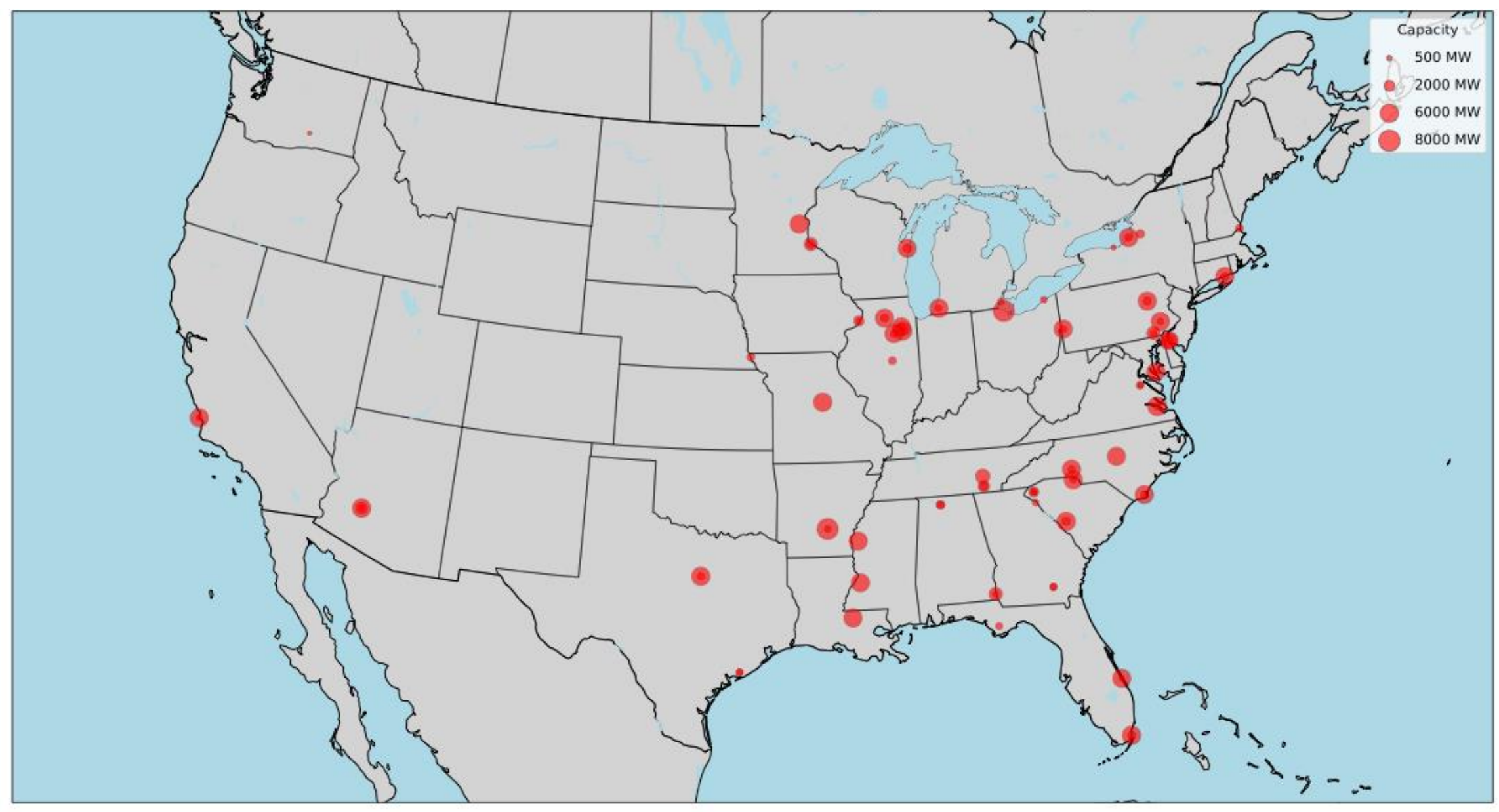

**Figure 4: Currently operational nuclear fission in power plants in the US. Source: EIA [52]**

Figure 3 shows the current state of data center and power grid infrastructure in the United States. We see that most of the data centers (existing, proposed, or under construction) are located close to major urban areas, and are heavily concentrated in the Midwest, east Coast and Texas. Given that these are already dense population centers, data center load growth will create significant stress on the local grid. For instance, Northern Virginia has one of the highest density of data centers in the US in an area dubbed "Data center alley" and this is one of the most congested grid networks in the country. There is also public opposition to building new transmission projects (e.g., power lines, substations) in such areas, making it very difficult to build and interconnected new generation. Although these areas often have solid fiber optic capacity for communication, grid transmission capacity is limited, especially when it comes to high voltage lines. This further motivates the need for co-located nuclear and data center solutions to reduce grid dependency. Comparing this with Figure 4, we see some interesting similarities between the locations of nuclear sites and data centers, especially in the mid-Atlantic and along the eastern seaboard – this complementary relationship could potentially be leveraged to our advantage. Given the preference of data center hyperscalers to site data centers close to cities, developing new fission could potentially be easier, faster and more streamlined compared to fission, due to its relatively lower safety concerns and regulatory hurdles. The limited availability and high cost of land in these areas also makes nuclear a more sensible option due to its higher energy and power density and lower land use requirements compared to renewables like wind and solar farms.

There are broadly two possible scenarios for deploying nuclear (either fission or fusion) to power data centers. Firstly, larger centralized nuclear power plants could be connected to the transmission grid and used to supply multiple data centers. This is more in-line with the current practices since the majority of data centers today rely primarily on grid connection along with some onsite backup power. A second approach would be to build data centers co-located with the nuclear power plant [53]. This is an emerging alternative to grid connection– e.g., hyperscalers have already started constructing large, combined cycle gas power plants co-located with data centers, which has disastrous consequences for carbon emissions. While connectivity to the main grid offers higher reliability, transmission grids are severely congested and capacity-constrained in most regions of the US [54]. This can delay or prevent the connection of large new generation plants. Upgrading, retrofitting, or building new grid infrastructure (including transformers, transmission power lines, etc.) is a time and capital-intensive process in the US. Such projects also often face regulatory hurdles, long permitting times, and local community opposition [55]. Figure 5 shows the interconnection queues for different ISOs in the US. We notice significant queues for all regions and across different project types, with a particularly long backlog for solar, battery storage, and wind projects [56]. Long interconnection processes hinder the integration of critical new renewables and storage projects. In addition to causing delays, they can sometimes even prevent new projects from getting off the ground altogether. This can be a serious challenge if we are to power new data centers with zero-carbon power. In fact, we observe that ISOs with long queues also have high expected demand growth from data centers [3]. To illustrate this, Figure 6 shows the cumulative interconnection queue at the end of 2023 for three independent system operators (ISOs) in the US, along with their respective projected datacenter demand growth by 2030. These three regions were chosen since these grid operators provide estimates for data center demand by 2030.

On the other hand, co-locating nuclear with data centers can reduce the grid capacity needed or potentially even eliminate the need for interconnection, if the site can be operated in an entirely islanded mode. Partial grid independence with nuclear power also offers some potential resilience gains without having to rely on expensive battery storage or auxiliary backup generation. Recently, there has also been an increased push from the US Department of Energy to co-locate data centers with new energy infrastructure on federal lands [57], as well as proposals for brownfield projects that repurpose retired coal [58] or nuclear plants [59] for data centers. Fusion plants also have more flexibility in location since they have fewer screening criteria and safety requirements that restrict where they can be installed, in comparison to fission plants [60]. Furthermore, it is relatively easier to 'right-size' the capacity of fusion plants based on the electricity needs of the co-located data centers. Data centers vary greatly in their capacities, ranging 1-5 MW for small, 5-20 MW for medium, and 20-1000+ MW for large sites [53]. These are generally smaller than the capacities of most fission reactors, which usually exceed 600 MW. While newer fission concepts like medium-scale (300-600 MW), small (50-300MW) and micro-modular (<50 MW) reactors are more flexible in sizing, these technologies are still largely untested and have not been deployed in the field. Thus, while fission plants can be well-suited for very large, hyperscale data centers, fusion may be a more flexible option for average-size data centers.

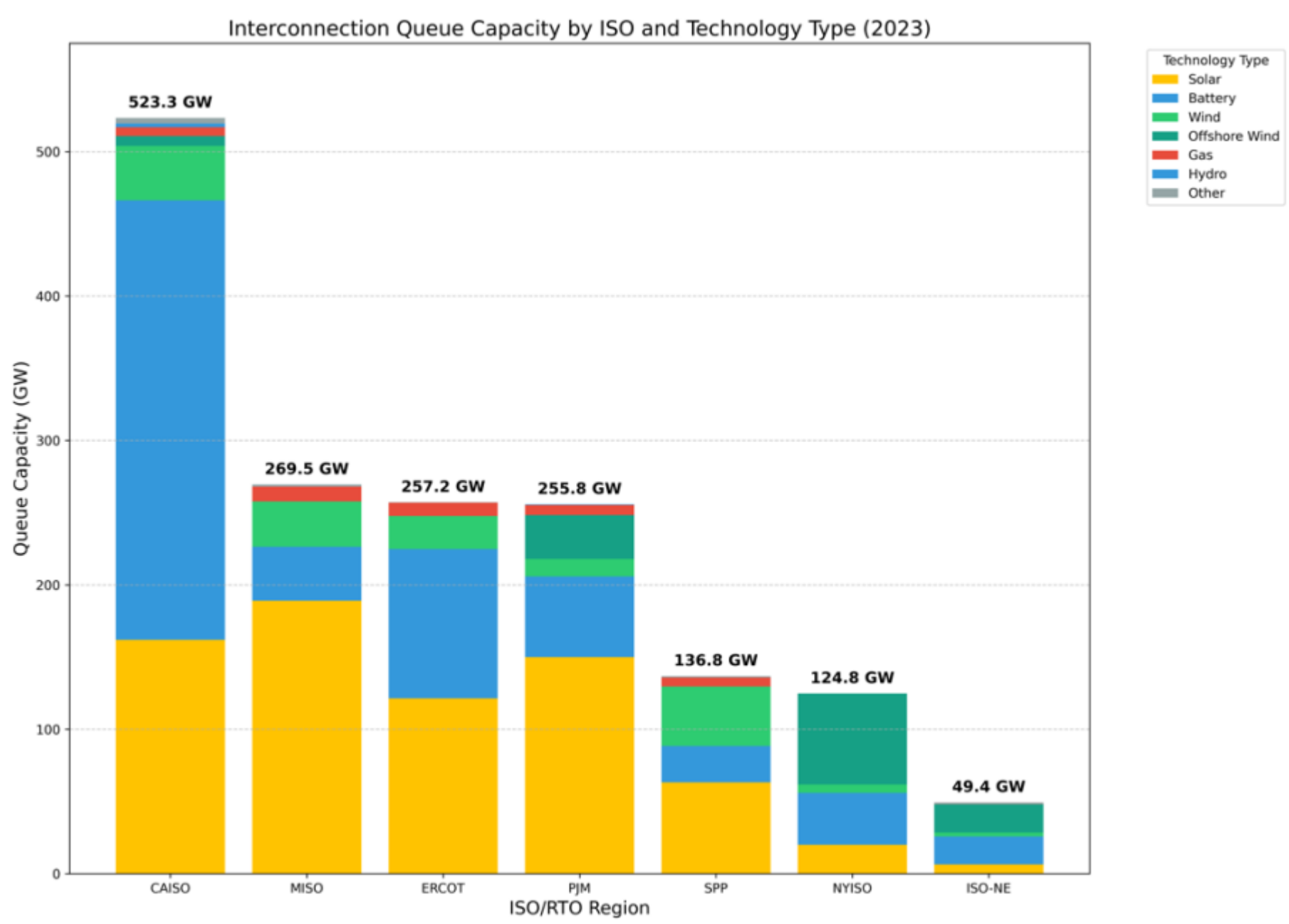

**Figure 5: Interconnection queues as of 2023 end for each ISO in the US.**

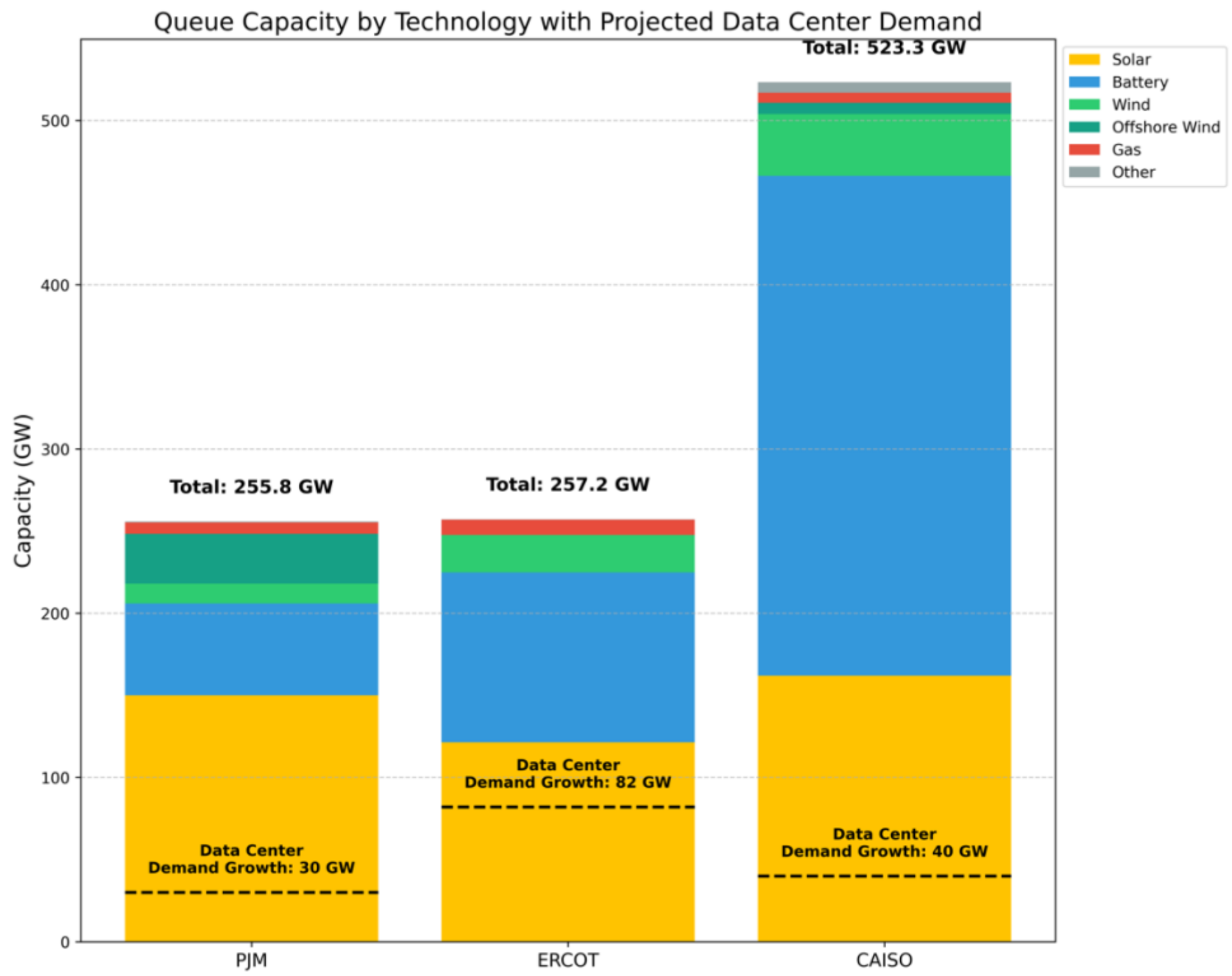

**Figure 6: Interconnection queue in 2023 vs projected data center demand growth by 2030.**

Finally, Figure 7 shows the growth of the interconnection queue between 2007-2023 across major US ISOs.

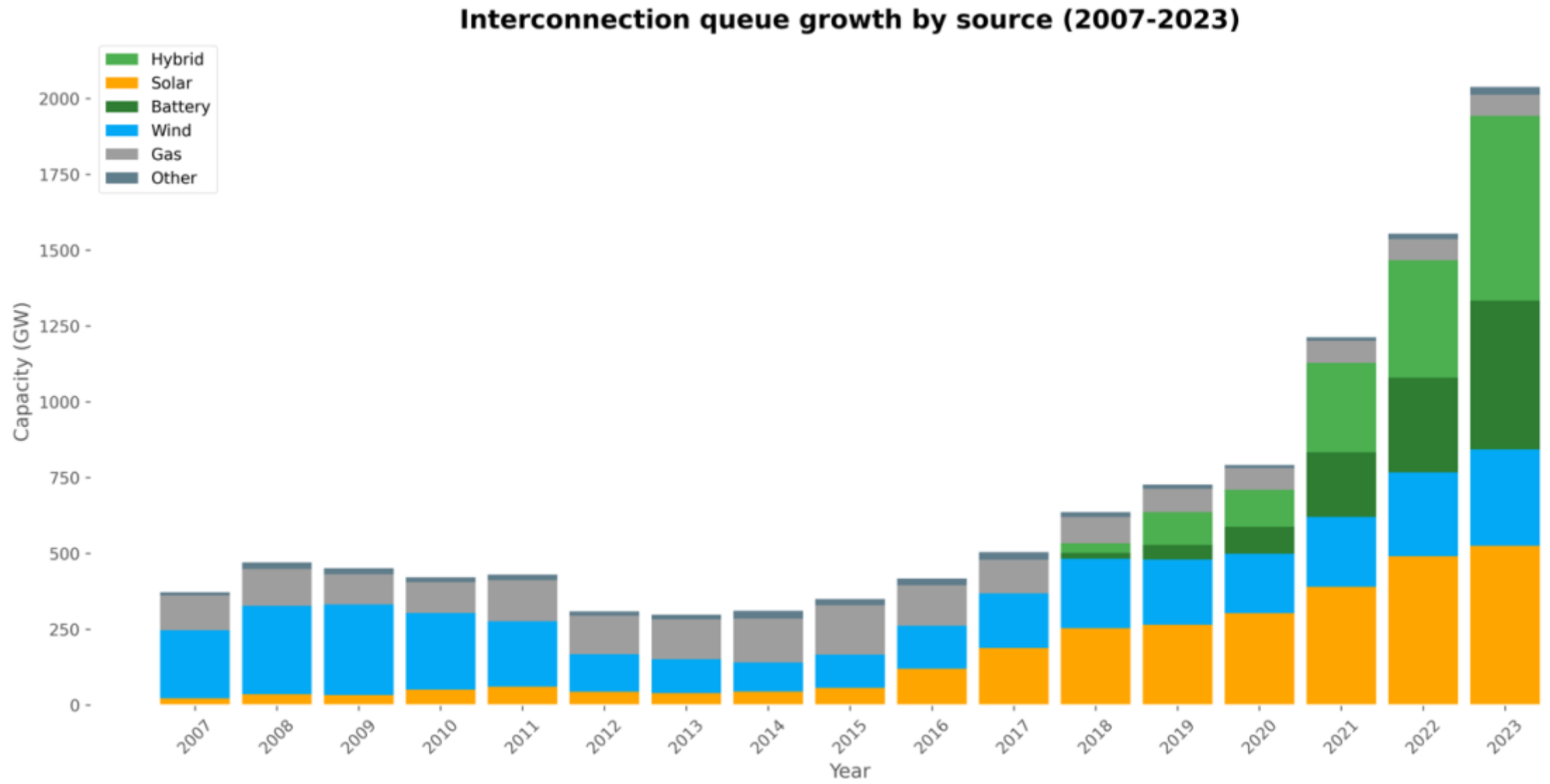

**Figure 7: Interconnection queue growth.**

## Holistic Assessment and Environmental Considerations

Regarding lifecycle emissions, Fusion's largest contributor to its carbon footprint is its replaceable components, for instance in DEMO, a tokamak power plant, the largest replaceable component is the blanket. It's need for replacement is due to neutronic damage [61]**.** Albeit DEMO is a FOAK reactor design, so as fusion matures its carbon footprint may shrink as durability of reactor components is realized. Nuclear has strong bipartisan support in the US, making it more resilient to changing political priorities, unlike e.g., wind power which has faced serious setback over the past year with the second Trump administration. Fusion is expected have lifecycle carbon emissions comparable to, or even lower than (for advanced NOAK reactors) fission.

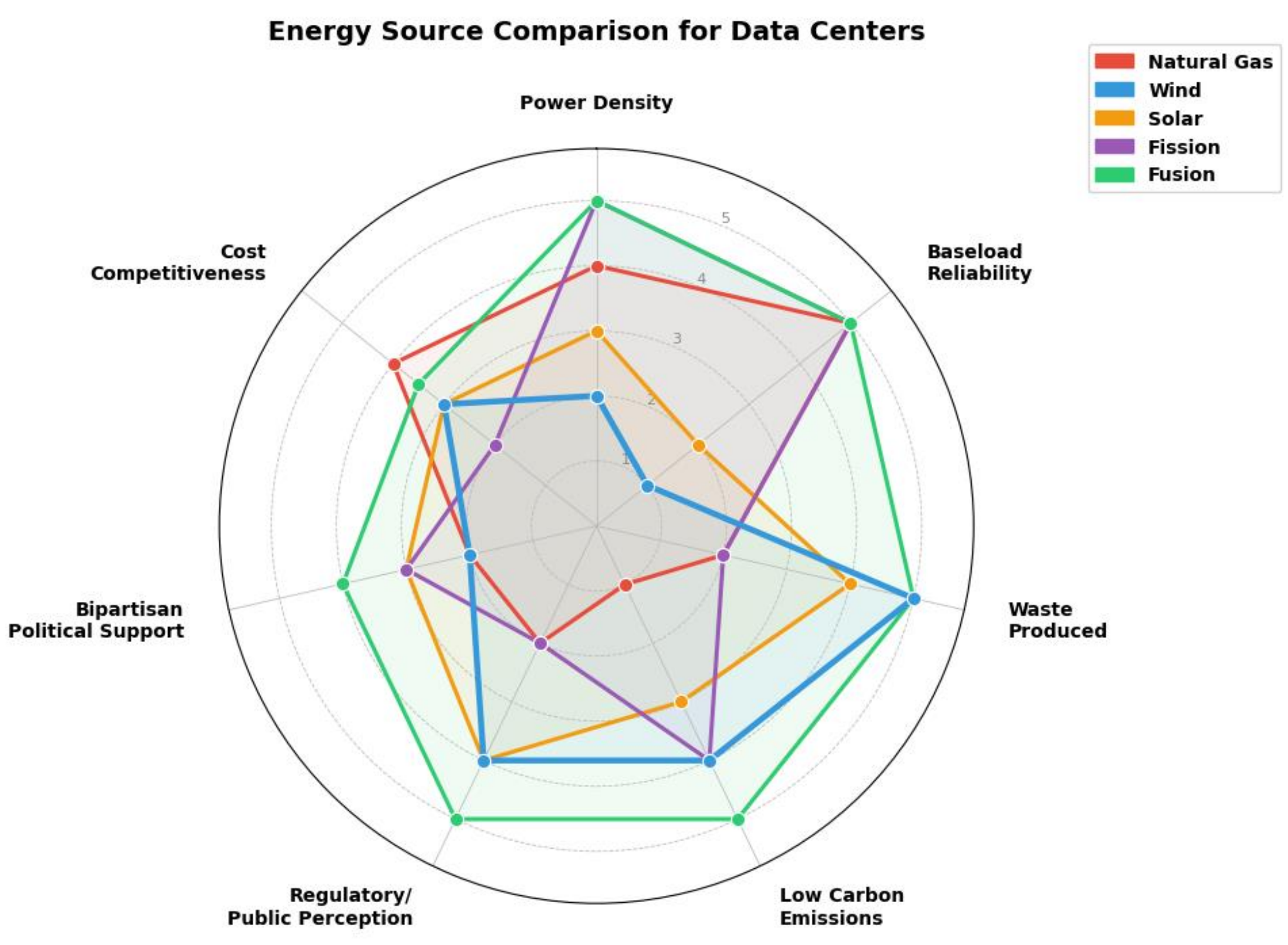

**Figure 8: Multi-Attribute Comparison of Power Generation Technologies for Data Centers.**

Figure 8 reflects a qualitative, system-level comparison across seven attributes relevant to data center deployment: power density, baseload reliability, waste produced, lifecycle carbon emissions, regulatory/public perception, and bipartisan political support. Fusion scores highly on power density and baseload reliability due to its continuous operation and compact energy footprint, comparable to fission and materially higher than wind and solar designs [58][62]and component replacement cycles [59]. Waste is treated more favorably than fission due to the absence of long-lived transuranic waste, though not negligible given activated structural materials [63]. Renewables score well on carbon emissions and public perception but are constrained by low power density and lack of dispatchability, which reduces their standalone suitability for high-load, always-on applications like data centers. Natural gas performs strongly on reliability and political acceptance but scores poorly on emissions. Fission maintains high marks for reliability and density but is penalized on waste and public perception. It's also important to note that there are emerging end-of-life waste management challenges associated with widespread deployment of renewables and batteries as well. Although these are less serious than nuclear waste concerns, there remain open issues around the proper reuse, recycling, and disposal of wind turbine blades [64], solar photovoltaic modules [65], and spent lithium-ion batteries [66].

Cost is scored on the inverse of unsubsidized levelized cost of energy, with variable renewables shown on a firmed basis to reflect the cost of supplemental capacity required to deliver dispatchable power equivalent to nameplate output. Firmed wind ranges from $60 to $153 per MWh and firmed solar ranges from $68 to $203 per MWh per Lazard's 2024 LCOE+ analysis, placing both in the same band as fission new build at $142 to $222 per MWh. Gas combined cycle remains the lowest cost firm option in the near term at $45 to $108 per MWh and does not require a firming adder. Under NOAK assumptions, Magnetic Confinement Fusion is estimated at $140/MWh, while Magneto-Inertial Fusion ranges from $50.8 to $85.5/MWh after adjustment to current dollars, overlapping the cost range of natural gas combined cycle and falling below the upper-bound costs of firmed wind and solar. On a firmed basis, the cost advantage of variable renewables narrows for always-on loads, while fusion spans a wider cost range that overlaps natural gas combined cycle at the lower end and remains competitive with firmed wind and solar at higher estimates. This cost position, combined with fusion's high scores for power density, reliability, emissions, and waste, strengthens its potential suitability for hyperscale data center loads.

Bipartisan support does not track the technical attributes shown in the figure and instead acts as a separate constraint on deployment. Recent U.S. policy signals have been uneven across technologies. Wind development has faced increasing federal friction through leasing pauses, expanded permitting reviews, and project-level interventions, including actions affecting projects already under construction[67]. Approval timelines have lengthened and become less predictable, with some projects delayed or reconsidered during review. These factors have reduced near-term pipeline visibility despite declining costs. Solar projects are experiencing a similar phenomenon to wind [67]. Natural gas remains operationally well suited for firm capacity but faces tightening policy constraints in several jurisdictions. Federal rules now require high levels of emissions control for new plants, and approvals for LNG export infrastructure have been paused[68]. At the state level, electrification mandates and decarbonization targets have limited the role of new gas infrastructure in regions with growing demand. The result is a reduced set of locations where gas expansion is both feasible and durable under current policy trajectories. This combination narrows the expansion pathway for both intermittent renewables and fossil baseload generation.

In contrast, nuclear energy has re-emerged as one of the few areas with sustained bipartisan alignment, driven by its ability to provide firm, low-carbon power while supporting domestic industrial policy and energy security goals. Fusion benefits further from its positioning as a forward-looking technology, with fewer legacy concerns around waste and safety, which moderates public opposition relative to fission. As a result, its higher score on bipartisan support reflects not just current policy, but relative durability under shifting political conditions[69]. Fusion's intermediate-to-high scores across all categories reflect its positioning as a hybrid: it inherits the firm, dense power characteristics of fission while avoiding some of its long-term waste and political liabilities, and it delivers low-carbon energy without the spatial and intermittency constraints of renewables. The plot is not intended as a precise quantitative ranking, but as a directional comparison to illustrate tradeoffs that are most relevant for siting and scaling large energy loads.

## Vision for co-development of AI and nuclear fusion

We believe there should be two different pathways we should focus on as we consider a timeline for fusion development. Firstly, efforts at deploying small- to medium-scale fusion plants, say by 2040, co-located with data center sites. These data center customers like hyperscaler tech companies, cloud computing, maybe even other industries that use AI can be the earlier adopter because they're willing to take on risk, have higher willingness to pay [70]. These also tend to have more funding available and can procure capital at relatively lower costs, particularly given the current strong emphasis and investment in AI across industries. This is crucial because high capital and financing costs lead to budget overruns and have been a major barrier to widespread fission adoption, especially in the US. Secondly, In the meanwhile, there should also be separate these stakeholders should aim to fund and support advanced fusion research and pilot or field-testing efforts. Both these efforts would bring down the levelized costs of nuclear power in general (both fission and fusion) in the US, which are much higher than in other countries like India and China [71]. We could thus become more competitive in the nuclear space, which is also important for energy independence and national security.

There have been several announcements recently where large tech companies (including Microsoft, Meta, Alphabet, and Amazon) have committed to either new nuclear fission projects or recommissioning existing plants. One specific example is the proposed reopening of the Three Mile Island plant to power Microsoft data centers. However, all of these involve conventional nuclear power plant designs. We believe there is a potential and need to also develop projects involving advanced fission reactors (such as small modular reactors) and fusion. In addition to nuclear, we have also seen recent examples of tech companies supporting, and investing, in more nascent and experimental technologies like advanced geothermal [72], thermal energy storage and thermophotovoltaics. Thus, fusion could potentially be another example of such collaborations and partnerships.

Thus, new focused policies and regulations must deal with both the public and private sector, while also recognizing the changing national strategy with the new administration in the US. However, the development of sustainable data centers also supports arguments for energy security, which is in line with both current and former US attitudes toward decarbonization. The other side of the issue is private industry which is still very motivated to reach net-zero or even net-negative operations (including scope 1, 2, and 3 emissions) with aggressive timelines. This necessitates the pursuit of reliable clean energy for data centers, to complement rapid AI growth.

## Conclusions

In this perspective, we argue that nuclear power (both fission and fusion) is uniquely well-suited to meet the rapid growth in power demand from AI and computing, particularly due to its firm baseload characteristics, dispatchability, and zero carbon footprint. The intermittency and variability of renewables like wind and solar creates challenges while serving data centers with high availability needs and require large amounts of grid-scale storage. Nuclear power offers much higher capacity factors and reliability, but lag behind in terms of costs, with significant uncertainty in projections. Fusion power plants are a promising alternative to fission in terms of regulatory complexity, social acceptance, and safety. However, realizing fusion's potential for data center applications will require reducing FOAK capital costs, establishing a credible pathway to cost-competitive NOAK deployment, and achieving plant availability consistent with the continuous power demands of hyperscale computing facilities. Nuclear's high density (in terms of both power and land use) makes it well-suited for co-location with data centers, which would reduce grid congestion and challenges with building or upgrading transmission infrastructure. We highlight the strong alignment between the needs of data centers and technical characteristics of fission and fusion power, particularly for AI training and inference. Through this perspective, we make an evidence-based argument that we must explore the potential of nuclear power, and fusion in particular, to sustainably meet computing load. We provide a balanced analysis that outlines their strengths but also discuss key important challenges and uncertainties that must be considered. We hope this will guide other researchers, practitioners and policymakers working in the nuclear, data center, and AI domains – to make widespread nuclear-power compute a reality.

## Acknowledgements

This research received no external funding. We would like to thank for Prof. George Tynan for helpful comments and feedback.